# Direct Wafer Bonding of Crystal-Ion-Sliced GaP Thin Films for Photonic Applications

Hossein Esfandiar [1,5,*], Emanuel Glück [1], Fabian Ganss [2], Ulrich Kentsch [2], Dennis Arslan[1], Jana Paeschke [1], Sebastian Ritter [1,5,6], Andreas Ihring [3], Muyi Yang [4,5,6], Isabelle Staude [4,5,6] Stefan Facsko [2], Falk Eilenberger [1, 5, 6], Carolin Rothhardt [1], Sebastian W. Schmitt [1, 5, +]

[1] Fraunhofer-Institute for Applied Optics and Precision Engineering IOF, Albert-Einstein-Str. 7, 07745 Jena, Germany

[2] Institute of Ion Beam Physics and Materials Research, Helmholtz-Zentrum Dresden-Rossendorf, Bautzner Landstr. 400, 01328 Dresden, Germany

[3] Leibniz Institute of Photonic Technology, Albert-Einstein-Str. 9, 07745 Jena, Germany

[4] Institute of Solid-State Physics, Friedrich Schiller University, Max-Wien-Platz 1, 07743 Jena, Germany

[5] Institute of Applied Physics, Abbe Center of Photonics, Friedrich Schiller University, Albert-Einstein-Str. 15, 07745 Jena, Germany

[6] Max Planck School of Photonics, Albert-Einstein-Str. 15, 07745 Jena, Germany

[*]hossein.esfandiar@iof.fraunhofer.de

[+]sebastian.wolfgang.schmitt@iof.fraunhofer.de

## Abstract

Gallium phosphide (GaP) is a promising material platform for integrated photonics because of its high refractive index, broad optical transparency, and strong second-order nonlinear response. Here, we demonstrate GaP-on-insulator thin films fabricated by crystal ion slicing and direct wafer bonding, using fused silica and $SiO_2$/Si thermally oxidized silicon substrates as representative platforms. Unlike GaP thin-film platforms that rely on heteroepitaxial growth or sacrificial-layer release, the presented approach enables the flexible integration of crystalline GaP thin films, independent of both donor and target substrates. Following post-transfer annealing, the films exhibit near-bulk crystalline quality with low residual strain, smooth surfaces suitable for nanophotonic fabrication, and homogeneous bonding interfaces. Furthermore, annealing restores the linear optical dispersion ($n$ and $k$) approaching that of epitaxially grown GaP with estimated plane wave absorption loss of 0.9 dB/cm at 1550 nm in the telecom C-band. The demonstrated approach establishes a scalable pathway toward high-quality GaP thin-film photonics compatible with versatile heterogeneous integration and back-end-of-line CMOS processing.

## Introduction

Among available materials for integrated nonlinear and quantum photonics, gallium phosphide (GaP) has attracted increasing attention due to its unique combination of properties [1]. GaP's refractive index exceeds 3 in the visible and near-infrared spectral range, enabling strong optical confinement in compact photonic structures [2]. Its indirect bandgap of 2.26 eV provides broad transparency across much of the visible spectrum while supporting strong second- and third-order nonlinear optical effects [3]. These properties make GaP highly attractive for integrated nonlinear optics, visible photonics, and quantum photonic devices [2, 4-8].

Recent progress in GaP nanophotonics has enabled a broad range of photonic structures including photonic crystal cavities [8,9], metasurfaces [10,11], and integrated nonlinear optical and optomechanical devices [2, 12-15]. High-quality GaP nanophotonic devices have demonstrated strong light confinement, low optical losses, and efficient nonlinear interaction across the visible and near-infrared spectral range [2]. However, scalable integration of high-quality crystalline GaP thin films with insulating substrates remains a significant challenge [16]. Most of the demonstrated GaP photonic platforms rely on epitaxially grown GaP/AlGaP heterostructures fabricated by molecular beam epitaxy (MBE) [4, 17] or metal–organic chemical vapor deposition (MOCVD) [18], followed by selective sacrificial-layer removal and thin film transfer processes. While these approaches have enabled high-performance photonic devices, they intrinsically depend on predefined epitaxial thin film stacks and substrate-specific layer growth architectures, limiting substrate choice and heterogeneous integration strategies [16].

Direct wafer bonding has emerged as an attractive route for integrating GaP thin films with insulating substrates, particularly in combination with oxide intermediate layers [4]. Nevertheless, previously demonstrated bonded GaP photonic platforms still primarily rely on epitaxially grown thin film structures derived from sacrificial-layer transfer processes [19]. In contrast, crystal ion slicing (CIS) enables flexible transfer of crystalline thin films by decoupling the active GaP thin film from the original growth substrate. Although CIS is widely established in other semiconductor material systems, its application to GaP photonic integration remains comparatively unexplored [20, 21].

Here, we demonstrate the fabrication of high-quality GaP-on-insulator thin films using CIS and plasma-activated direct wafer bonding onto silica and $SiO_2$/Si substrates. The transferred films exhibit uniform, defect-minimized bonding interfaces and smooth ion-sliced surfaces. Following GaP film transfer and annealing, implantation-induced crystal damage is substantially reduced, restoring the films toward near-bulk crystalline quality with low residual strain. Their refractive indices closely match those of bulk GaP across the visible and near-infrared spectral ranges, with an estimated planar-wave absorption loss of only 0.9 dB/cm at 1550 nm in the telecom C-band. This approach establishes a scalable pathway toward low-loss, high-quality GaP thin-film photonics compatible with versatile heterogeneous integration and back-end-of-line CMOS processing.

**Results and discussion**

To obtain an ~800 nm-thick GaP thin film, a CIS process, as schematically illustrated in Fig. 1a, was employed. Two (100)-oriented, single-side polished GaP single-crystal substrates with dimensions of 1 × 1 cm² were prepared as donor samples. These samples were subjected to ion implantation of helium ($He^+$) and co-implantation of hydrogen/helium ($H^+$/$He^+$) species, respectively to create a buried damage layer that enables controlled exfoliation. $He^+$ implantation was carried out at a kinetic energy of 105 keV, while $H^+$ ions were implanted at 70 keV, maintaining a fixed fluence of $1 \times 10^{17}$ ions $cm^{-2}$ both for $He^+$ and $H^+$.

As depicted in Fig. 1a, the CIS sequence proceeds as follows (for details on the different steps please refer to the Methods section): Following ion implantation (1), a 700 nm silicon dioxide ($SiO_2$) interlayer is deposited onto the GaP surface (2) to facilitate subsequent bonding and to act as an adhesion and mechanical support layer. The implanted donor wafer is then directly bonded to an acceptor substrate using plasma-activated direct bonding (3), which consists of either a thermally oxidized silicon wafer with a 3 µm $SiO_2$ thin film (without any restriction of the principal generality of the proposed approach for any other sample with an $SiO_2$ surface layer) or a fused silica (glass) substrate, depending on the intended photonic platform. Prior to bonding, both the donor ($SiO_2$-coated implanted GaP) and acceptor substrates undergo rigorous surface preparation, including chemical mechanical polishing (CMP) and standardized wet cleaning protocols, to achieve the sub-nanometer surface roughness and high surface energy required for direct wafer bonding. Subsequent to the bonding, controlled thermal annealing in a conventional furnace in air, induces hydrogen- and helium-related blistering and crack propagation along the weakened buried layer, resulting in exfoliation and transfer of the GaP thin film onto the target substrate after approximately 3 h at 350 °C (4). Following thin film transfer, the sample undergoes a higher-temperature annealing at 600 °C in vacuum (4), which serves to reconstruct the GaP lattice, reduce implantation-induced point defects and dislocation complexes, and restore crystallinity toward bulk-like structural quality.

Figure 1b shows TRIDYN simulation results for the applied $He^+$ and $H^+$ implantation parameters. Asymmetric Gaussian fits indicate a projected ion range ($R_p$) centered at 844.5 nm and 806.5 nm beneath the GaP surface for He and H ions respectively, with a longitudinal straggle of ~200 nm full width at half maximum (FWHM). This depth distribution closely matches the targeted thin film thickness and ensures optimal energy deposition and defect accumulation within the intended cleaving plane.

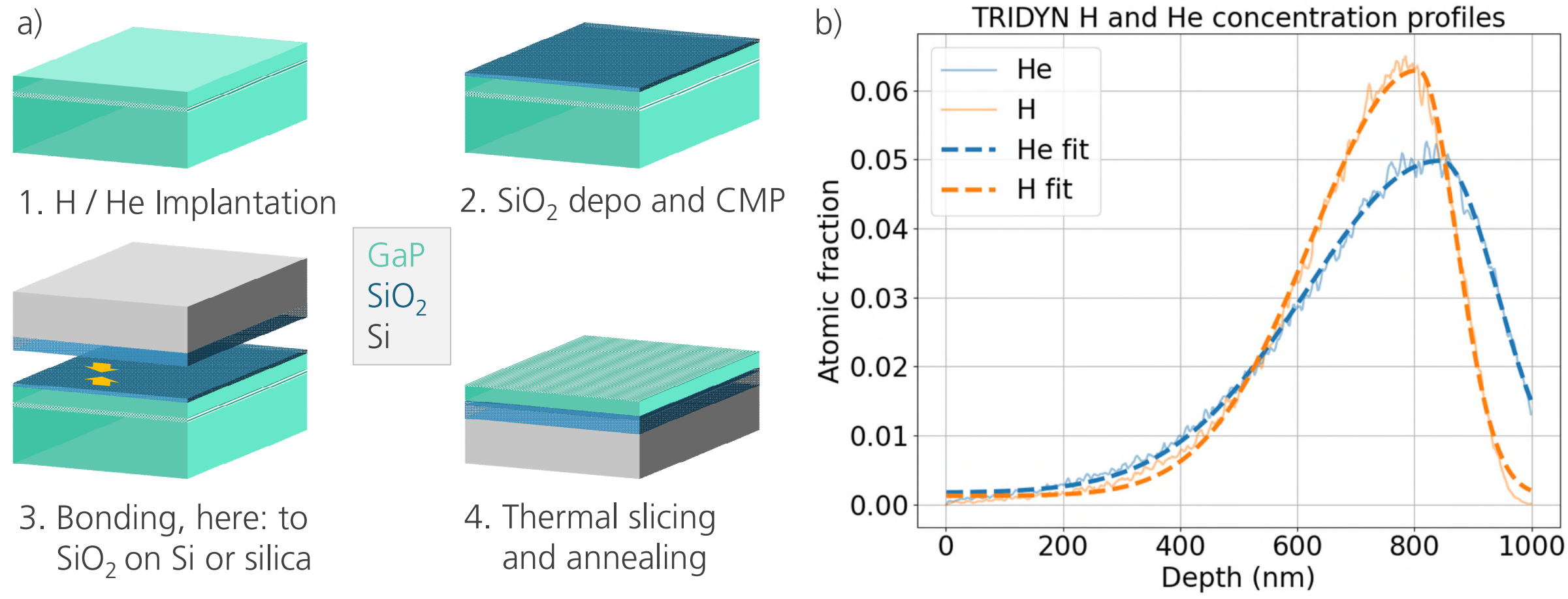

**Figure 1.** Fabrication process flow and projected ion range: **(a)** Process flow for crystal ion slicing and plasma activated direct bonding process. **(b)** TRIDYN-calculated depth distribution of 105 keV He ions and 70 keV H ions implanted into a GaP crystal at a fluence of $1 \times 10^{17}\,cm^{-2}$. The dashed lines show fits of the curves with split (asymmetric) Gaussian functions.

Following wafer bonding and thermal exfoliation, implanted GaP layers were successfully transferred onto the respective target substrates. The $He^+$-implanted donor wafer was bonded to a thermally oxidized $Si/SiO_2$ substrate, whereas the $H^+/He^+$ co-implanted donor wafer was bonded to fused silica. IR transmission and optical microscopy (OM) images of the two transferred layers are presented in Figs. 2a and 2b, respectively.

As shown in Fig. 2a, the IR transmission image of the $He^+$-implanted sample shows no extended interfacial voids within the bonded area. The corresponding OM image, however, reveals a discontinuous GaP layer containing numerous circular regions in which material was not transferred, resulting in a porous appearance. The localized nature of these defects could arise from several factors, including surface roughness, particulate contamination, local variations in interfacial bonding, or spatial variations in crack initiation and propagation during thermal exfoliation. The available data does not allow the relative contributions of these possible mechanisms to be distinguished.

The $H^+/He^+$ co-implanted sample shown in Fig. 2b also exhibits an almost void-free bonded interface in the IR transmission image, apart from a single isolated defect marked by the yellow arrow. This defect may, for example, have originated from a dust particle or another localized surface imperfection. In contrast to the $He^+$-implanted sample, the corresponding OM image shows a largely continuous and homogeneous GaP layer across the target substrate.

These observations demonstrate that GaP layer transfer was achieved using both implantation and bonding configurations, although the resulting films exhibited substantially different degrees of continuity. Because the two experiments differed in both implantation conditions and target substrate, the improved continuity of the $H^+/He^+$ co-implanted sample cannot be attributed uniquely to the co-implantation process. Substrate-dependent surface properties, surface preparation, local contamination, and differences in interfacial bonding may also have contributed to the observed transfer morphologies. A systematic comparison involving repeated experiments on identical target substrates would be required to separate these effects and establish the influence of the implantation conditions conclusively.

Nevertheless, $H^+/He^+$ co-implantation provides a physically plausible mechanism for facilitating controlled layer separation. During thermal annealing, implanted hydrogen can promote the formation and growth of $H_2$-containing platelets, while helium can interact with implantation-induced vacancies and contribute to the stabilization and pressurization of subsurface defect clusters. Their combined action may facilitate the formation of a buried fracture plane and promote layer separation. The comparatively high continuity of the co-implanted film observed here is consistent with this mechanism, although the present results do not demonstrate that co-implantation was the sole or dominant cause of the difference between the two samples.

Taken together, the results establish proof of concept for GaP thin-film transfer using both $He^+$ implantation and $H^+/He^+$ co-implantation under the respective processing conditions. Among the specimens investigated, the $H^+/He^+$ co-implanted GaP layer transferred to fused silica exhibited the greater film continuity and was therefore selected for the subsequent analyses. This selection is based on the quality of the obtained specimen and should not be interpreted as a general comparison of the two implantation schemes or target substrates. However, where available, complementary characterization results for GaP films bonded to Si/$SiO_2$ are provided in the Supporting Information.

Figure 2c shows a cross-sectional SEM image of the GaP thin film on fused silica. Within the inspected cross section, the interfaces between the GaP film and the bonding $SiO_2$ layer, as well as between the $SiO_2$ layer and the fused silica substrate, appear continuous and free of resolvable voids. The measured GaP film thickness is approximately 790 nm, in good agreement with the projected ion ranges of $H^+$ and $He^+$ obtained from the simulations shown in Fig. 1b. The $SiO_2$ layer has a measured thickness of approximately 700 nm, compared with the nominal thickness of 700 nm after deposition and approximately 670 nm expected after the subsequent chemical–mechanical polishing process.

Surface topography remaining after ion slicing and transfer was visible in the SEM image and was examined in greater detail by atomic force microscopy (AFM). The AFM measurement shown in Fig. 2d was performed over an area of 10 × 10 µm² and provides a local assessment of the transferred GaP surface. An RMS roughness of approximately 4.9 nm was obtained, indicating nanoscale surface roughness within the investigated area. This result provides a reference for evaluating the need for further surface treatment before subsequent microfabrication or device-processing steps.

Figure 2e shows a photograph of the GaP thin film bonded to fused silica after annealing. The relatively homogeneous macroscopic appearance and optical transmission across most of the film are consistent with the IR transmission and OM observations in Fig. 2b. The position of the isolated defect marked by the yellow arrow corresponds to the defect location identified in the IR transmission image.

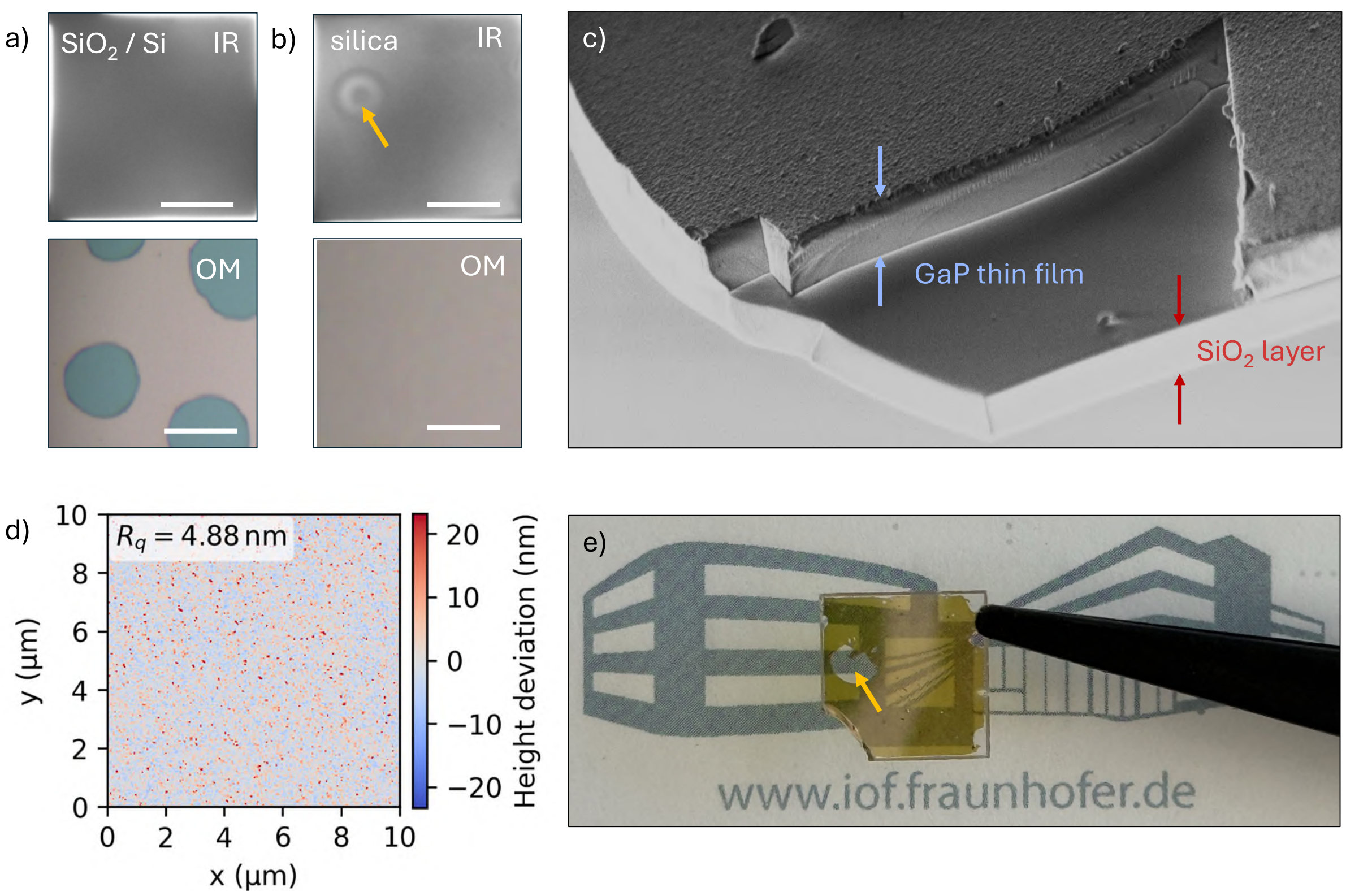

**Figure 2.** Surface and bonding interface characterization: IR transmission (top) and optical microscopy of surface area (bottom) of **(a)** $He^+$ ion-sliced GaP thin film bonded on $SiO_2$ on Si and **(b)** $He^+/H^+$ ion sliced GaP thin film bonded on silica. The scale bars are 3 mm (top) and 8 µm (bottom), respectively. **(c)** SEM tilt view of the cross section of the layer stack. The GaP thin film and $SiO_2$ layer thickness indicated by the arrows are 790 nm and 670nm, respectively. **(d)** AFM surface topography and **(e)** photograph of the $He^+/H^+$ ion sliced GaP thin film on silica (after annealing).

Figure 3 shows the (224) reciprocal space maps (RSMs) of the pristine bulk GaP (Fig. 3a), the GaP thin film following crystal $He^+/H^+$ ion slicing and bonding to a silica substrate (Fig. 3b), and the same bonded GaP thin film after annealing at 600 °C for 12 h in vacuum (Fig. 3c). In each map, the black open circle denotes the peak position obtained from a two-dimensional Gaussian fit of the data while the black cross indicates the reciprocal lattice coordinates of the bulk pristine GaP ($Q_x/(2\pi)$ = 5.1917 nm$^{-1}$, $Q_z/(2\pi)$ = 7.3380 nm$^{-1}$). Accordingly, the cross coincides with the fitted peak center in Fig. 3a, while the offsets between circle and cross observed in Figs. 3b and 3c indicate deviations from the pristine bulk GaP lattice.

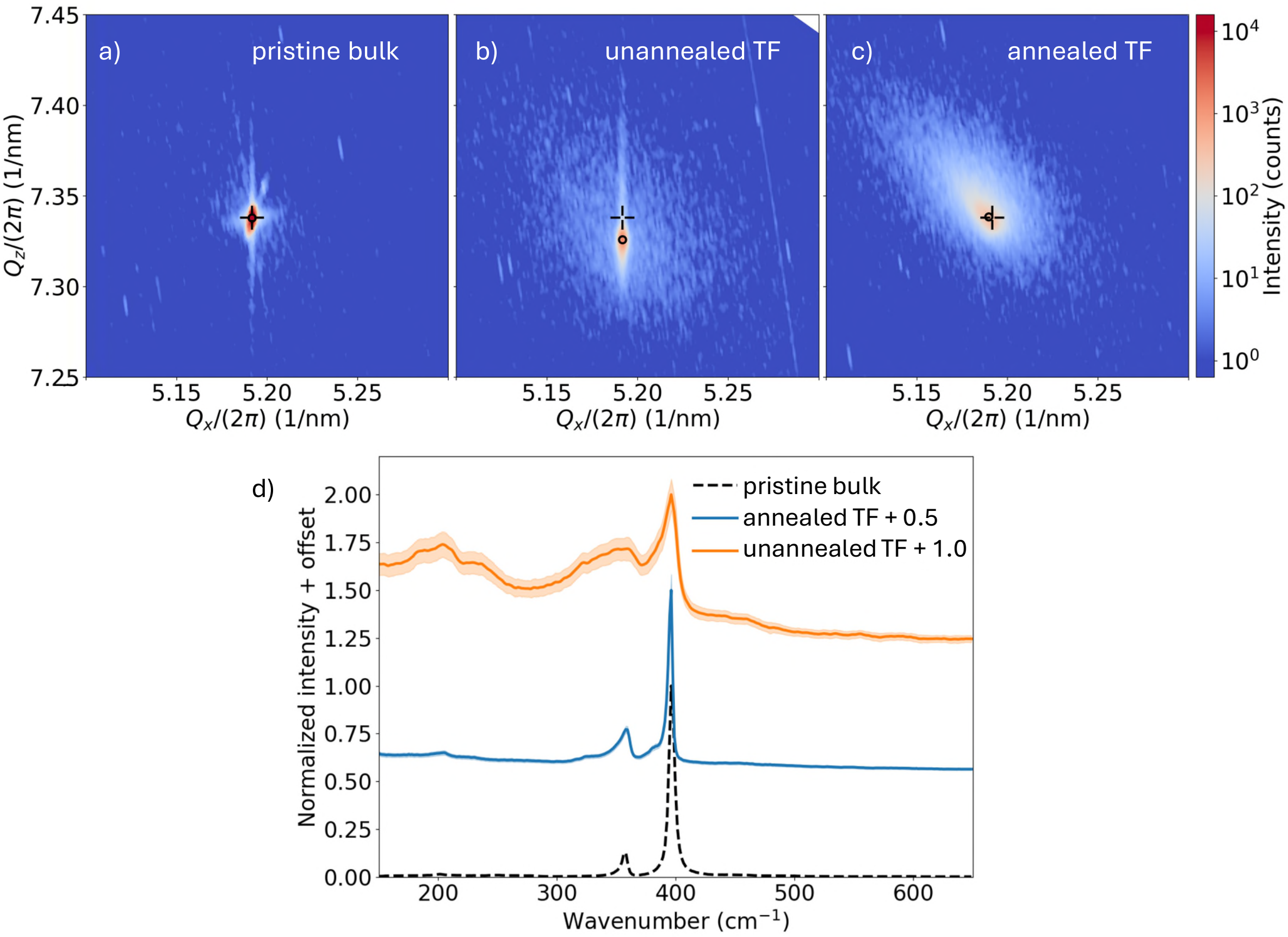

**Figure 3.** Structural investigation of the GaP before and after exfoliation, bonding and annealing. Reciprocal space maps (with logarithmic intensity scale) of GaP around the (224) reflection for **(a)** pristine bulk GaP, **(b)** $He^+$ / $H^+$ co-implanted, exfoliated and bonded GaP thin films, and **(c)** the same sample after annealing for 12 h at 600°C in vacuum. **(d)** Raman spectra of bulk GaP, the $He^+/H^+$ ion sliced and bonded GaP film before annealing, and the film after annealing at 600 °C for 12 h in vacuum.

The pristine bulk GaP crystal (Fig. 3a) exhibits a sharp, diffraction peak with minimal diffuse scattering, characteristic of a high-quality single crystal. The narrow-fitted widths of FWHM($Q_x$)/(2π) = 1.63×10$^{-3}$ nm$^{-1}$ and FWHM($Q_z$)/(2π) = 5.05×10$^{-3}$ nm$^{-1}$ confirm the excellent crystalline quality and low defect density of the pristine bulk material.

Following crystal ion slicing and direct bonding to silica (Fig. 3b), the diffraction peak becomes noticeably broader and is accompanied by pronounced diffuse scattering extending around the Bragg reflection. The peak is elongated predominantly along the $Q_z/(2\pi)$ direction, indicating an increased distribution of the out-of-plane lattice parameter. The fitted peak remains at essentially the same in-plane reciprocal coordinate ($Q_x/(2\pi)$ = 5.1917 nm$^{-1}$) but shifts to a lower out-of-plane reciprocal coordinate ($Q_z/(2\pi) = 7.3259$ nm$^{-1}$). This decrease in $Q_z/(2\pi)$ clearly reflects an expansion of the out-of-plane lattice parameter, likely due to the presence of implanted ions. The peak widths increase to FWHM($Q_x$)/(2π) = 2.84×10$^{-3}$ nm$^{-1}$ and FWHM($Q_z$)/(2π) = 8.42×10$^{-3}$ nm$^{-1}$, demonstrating a significant implantation-induced strain gradient accompanied by reduced long-range order. The enhanced diffuse scattering further confirms the presence of implantation-induced defects introduced during the crystal ion slicing and bonding processes.

After annealing at 600 °C in vacuum for 12 h (Fig. 3c), the diffraction peak shifts back towards the pristine bulk GaP reciprocal lattice position, with the fitted peak located at $(Q_x,Q_z)/(2\pi) = (5.1897, 7.3385)$ nm$^{-1}$ indicating that the implantation-induced lattice strain has been largely relieved. Visually, the diffuse scattering surrounding the Bragg peak is merely changing shape compared with the as-bonded film, demonstrating a remaining reduced lattice uniformity. The diffraction peak appears more symmetric and centered on the nominal GaP reciprocal lattice position, although it is noticeably broader (FWHM($Q_x$)/(2π) = 1.14×10$^{-2}$ nm$^{-1}$, FWHM($Q_z$)/(2π) = 1.30×10$^{-2}$ nm$^{-1}$) than that of the pristine bulk GaP and the film before annealing likely indicating remaining strain gradients.

Figure 3d presents the Raman spectra of bulk GaP, the $He^+/H^+$ ion-sliced and bonded GaP film before annealing, and the film after annealing at 600 °C for 12 h in vacuum. The pristine GaP spectrum exhibits the characteristic transverse optical (TO) and longitudinal optical (LO) phonon modes of crystalline GaP [22]. Pseudo-Voigt fitting yields a TO linewidth of 5.44 ± 2.18 cm$^{-1}$ and an LO linewidth of 5.63 ± 0.27 cm$^{-1}$.

Following crystal ion slicing and bonding, the LO phonon remains clearly visible, confirming preservation of the crystalline GaP phase, but the Raman spectrum becomes significantly broadened and is superimposed on an elevated background, indicative of implantation-induced lattice disorder, strain inhomogeneity, and defect-assisted scattering. Fitting shows that the LO linewidth increases to 12.13 ± 0.88 cm$^{-1}$, consistent with substantial structural damage and increased phonon scattering. Although the fitting routine returned a TO peak at the upper fitting boundary (380 cm$^{-1}$) with the maximum allowed linewidth (30 cm$^{-1}$), this component does not represent a physically resolved TO phonon and instead reflects the broad disorder-related background. Consequently, the fitted TO linewidth for the unannealed sample is not considered physically meaningful and is excluded from the discussion.

After annealing at 600 °C for 12 h, the Raman spectrum exhibits a marked reduction in the disorder-related background together with significantly sharper phonon features. The LO linewidth decreases to 3.97 ± 0.46 cm$^{-1}$, demonstrating substantial recovery of crystalline order and a reduction in local strain fluctuations. The TO phonon is again clearly resolved at 357.12 ± 1.14 cm$^{-1}$ with a linewidth of 8.31 ± 3.14 cm$^{-1}$, indicating recovery of the lattice vibrations that were obscured in the as-transferred film.

Overall, the Raman results provide quantitative evidence that thermal annealing effectively reduces implantation-induced defects, restores well-defined optical phonon modes, and significantly improves the structural quality of the transferred GaP film. These observations are in good agreement with the reciprocal space mapping results, which likewise demonstrate relaxation of implantation-induced lattice strain and recovery of crystalline order after annealing.

Compared with the $He^+$-implanted film for which XRD and Raman results are shown in Supplementary Fig. S1 and the following discussion, the $He^+/H^+$ co-implanted film exhibits greater initial out-of-plane lattice expansion and stronger implantation-induced disorder, as indicated by its larger negative $\Delta Q_z$ and broader LO Raman linewidth before annealing. Following annealing, both films recover closely to the pristine bulk GaP lattice position. However, the annealed $He^+$-implanted film on $SiO_2$/Si shows a much narrower RSM peak width after annealing, suggesting lower residual strain inhomogeneity. Both films exhibit similarly narrow LO phonon linewidths after annealing, demonstrating substantial recovery of long-range crystalline order in either case.

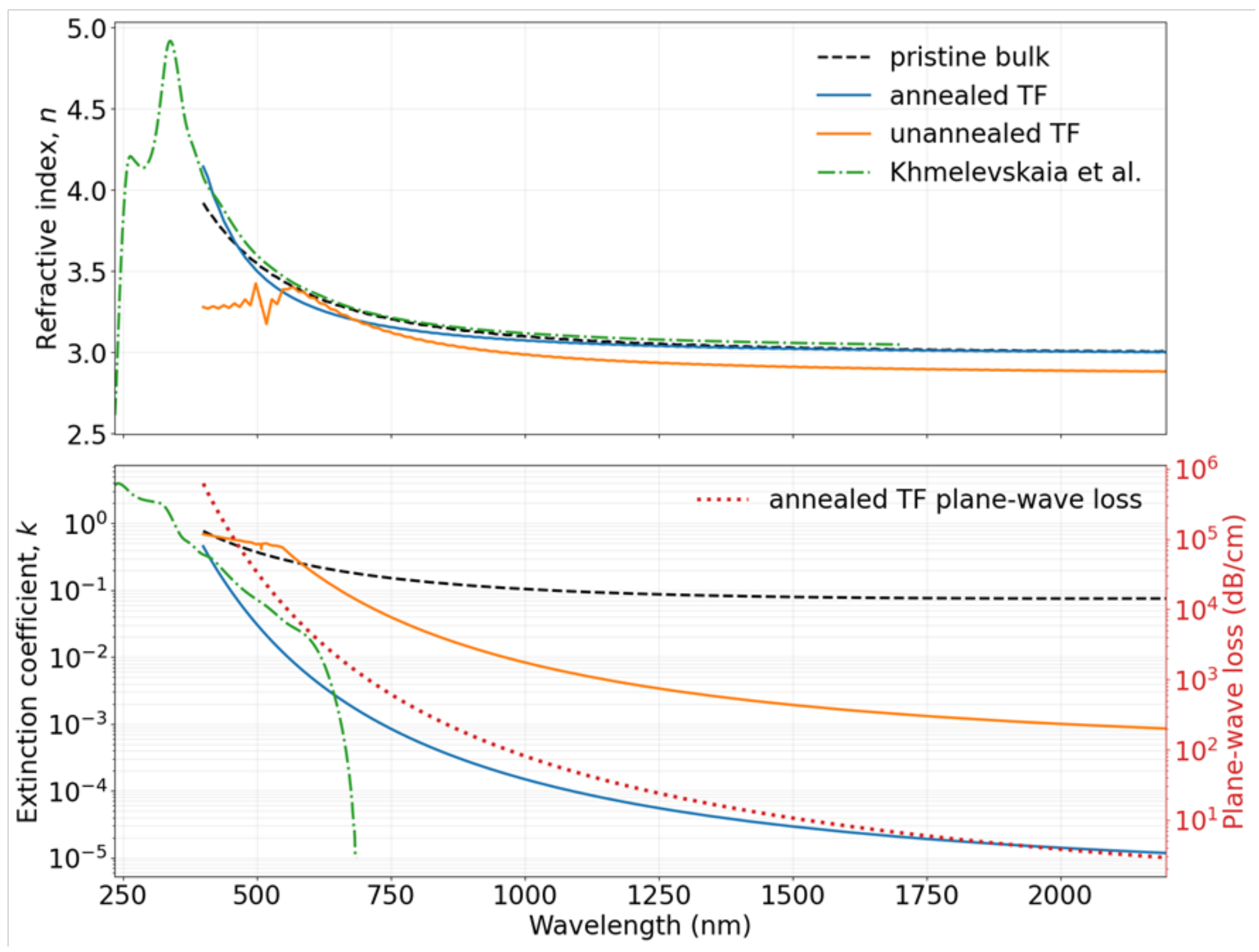

**Figure 4:** Spectral dependence of the complex refractive index of $He^+/H^+$ co-implanted, ion-sliced GaP thin films bonded to silica fused before and after annealing at 600 °C for 12 h. The upper and lower panels show the refractive index $n$ and extinction coefficient $k$, respectively. Results are compared with the pristine bulk GaP and literature data for crystalline GaP grown directly on sapphire [17]. The red dotted line lower panel shows the wavelength dependent plane wave propagation loss in the annealed thin film.

Figure 4 compares the complex refractive index, ($N=n+ik$), of the pristine bulk GaP, the $He^+/H^+$ co-implanted and bonded GaP thin film before and after annealing, and the directly grown crystalline GaP-on-sapphire film reported by Khmelevskaia et al. [17]. The unannealed transferred film exhibits a systematically lower refractive index than the pristine bulk GaP, particularly in the near-infrared, together with an elevated extinction coefficient. These differences are attributed to implantation-induced defects, strain inhomogeneity, and reduced lattice order, which modify the dielectric response and introduce sub-bandgap optical absorption.

After annealing at 600 °C for 12 h, the refractive-index $n$ of the thin film shifts toward that of bulk GaP and becomes nearly identical to the pristine bulk GaP at longer wavelengths. More importantly, the extinction coefficient decreases by approximately one to two orders of magnitude relative to the unannealed film throughout much of the near infrared spectral range. This strong reduction in $k$ demonstrates the removal or passivation of optically active implantation defects and a corresponding decrease in sub-bandgap optical loss. The spectral irregularities observed around 500–600 nm occur close to the GaP absorption edge, where the optical constants are strongly dispersive and the ellipsometry fitting is more sensitive to the assumed film thickness, surface roughness, and interface model.

The pristine bulk GaP sample measured here exhibits an anomalously large extinction coefficient at long wavelengths, suggesting appreciable defect-related absorption and lower optical quality than expected for high-quality bulk GaP. However, because small $k$ values are particularly sensitive to measurement uncertainty and ellipsometric model assumptions, contributions from fitting artifacts, surface roughness, or backside reflections cannot be excluded. The measured sample should therefore not be regarded as representative of the intrinsic absorption limit of high-quality bulk GaP. The directly grown GaP-on-sapphire film reported by Khmelevskaia et al. also exhibits strong band-to-band absorption at short wavelengths, although its extinction coefficient rapidly decreases toward the transparent spectral region. In comparison, the annealed ion-sliced GaP film combines bulk-like refractive-index dispersion with the lowest near-infrared extinction coefficient among the samples considered. Overall, the ellipsometry results demonstrate substantial optical recovery after annealing and are consistent with the RSM and Raman measurements, which independently reveal strain relaxation, defect reduction, and improved crystalline order in the transferred GaP film. The wavelength-dependent plane-wave propagation loss of the annealed thin film is shown as a red dotted line in the lower panel. Assuming complete optical confinement within the film, it represents an upper bound on the material-absorption contribution to the waveguide loss and reaches 0.9 dB/cm at the telecommunication wavelength of 1550 nm.

## Conclusions and Outlook

We have demonstrated an approach to fabricating high-quality GaP-on-insulator thin films by combining crystal ion slicing (CIS) with direct wafer bonding. This process enables crystalline GaP films to be transferred onto both silica and $SiO_2$/Si substrates, providing a flexible alternative to conventional platforms based on epitaxial growth and sacrificial-layer release. The transferred films exhibit smooth surfaces and homogeneous, defect-minimized bonding interfaces suitable for subsequent nanophotonic processing.

Thermal annealing substantially restores the structural and optical properties of GaP after ion implantation. Raman spectroscopy reveals the recovery of well-defined optical phonon modes, while X-ray diffraction and reciprocal-space mapping show improved crystalline order and reduced implantation-induced lattice strain. Consistently, the refractive index and extinction coefficient approaches those of crystalline GaP. The wavelength-dependent plane-wave propagation loss, calculated from the extinction coefficient of the annealed film, reaches 0.9 dB/cm at 1550 nm in the telecom C-band. Assuming complete optical confinement within the GaP film, this value represents an upper bound on the material-absorption contribution to the waveguide loss.

The wafer-scalable bonding process offers considerable flexibility for heterogeneous integration with established photonic platforms. Its compatibility with silicon-on-insulator substrates is particularly attractive for silicon photonics and CMOS processing, while avoiding epitaxial GaP growth. GaP's large bandgap and high refractive index also make the platform promising for visible-wavelength applications, enabling increased integration density and access to a broad range of light sources, quantum emitters, and high-quality electro-optic devices. This approach may therefore pave the way toward next-generation active and quantum photonic integrated circuits.

## Methods

### Material Preparation and Ion Implantation

A 2-inch diameter one-side polished GaP (100) [23] is diced into small sample sizes of $1\times1$ cm$^2$. The samples are implanted at room temperature with $He^+$ and co-implantation of $H^+$ /$He^+$ ions. The ion implantation was performed at room temperature using an air insulated 500 kV ion implanter with an indirectly heated cathode (IHC) ion source (Bernas type), a 40 kV extraction system and a 460 kV post acceleration unit. $He^+$ implantation was performed with an energy of 105 keV while $H^+$ was implanted by 70 keV at a fixed fluence of $1 \times 10^{17}$ He cm$^{-2}$. During implantation, the surface normal to the sample was tilted at an angle of 7° with respect to the ion beam direction in order to avoid channeling effects.

### X-ray Diffraction

Reciprocal space maps (RSM) were recorded on a Rigaku SmartLab 3 kW X-ray diffractometer with a Ge (220) two-bounce monochromator and a HyPix-3000 detector in 1D mode. The incident slit was set to 0.05 mm. Reciprocal space maps (RSMs) were acquired around the asymmetric GaP (224) reflection to evaluate the strain state, and crystalline quality of the transferred GaP thin films.

### Wafer Bonding

Plasma-activated direct bonding (PADB) was used to join the substrates through direct interaction of their surface atoms, eliminating the need for organic adhesive layers that can compromise optical performance. At the same time, the method preserves the optical properties of the materials while providing a mechanically robust interface. In combination with CIS, PADB therefore offers a promising route for the fabrication of GaP-on-insulator substrates. Because PADB relies on atomic-scale contact between the bonding surfaces, low surface roughness and flatness are required. The surface roughness over a 10 × 10 µm² area should remain below 1 nm root mean square, and the flatness of thin chips should be on the order of 1 µm peak-to-valley. Established models can be used to assess the bonding suitability of such surfaces prior to processing [24]. PADB is a mature technique in semiconductor fabrication and has been shown to provide highly reliable $SiO_2$–$SiO_2$ interfaces in both bulk and coated systems [25]. Owing to its low refractive index, $SiO_2$ is also well suited as an insulating layer for optical waveguide platforms.

Following deposition of a 700 nm $SiO_2$ layer on GaP, chemical mechanical polishing (CMP) was performed to obtain the required surface quality. AFM roughness measurements and Fizeau interferometry flatness measurements confirmed that the coated GaP substrate, as well as the oxidized silicon and fused silica substrates, fulfilled the requirements for direct bonding. Prior to bonding, the substrates were cleaned using ultrasonic bath-assisted procedures developed for direct bonding, including an adapted RCA spin-cleaning step. This was followed by low-pressure plasma activation using nitrogen and oxygen to enhance the bonding strength of the silicon dioxide interfaces. A subsequent annealing step at 150 °C induced condensation of the plasma-activated hydrophilic $SiO_2$ surfaces, resulting in the formation of covalent siloxane bonds. The annealing temperature was selected to avoid thermal splitting of the implanted layer while still promoting sufficient bond strength for stable adhesion of the crystalline layer. On the basis of previous measurements on coated samples, the bonding strength is expected to reach up to 50% of the bulk material strength.

**Thermal Processing and Layer Transfer**

Following the bonding process, the GaP crystals were annealed in a conventional furnace at 350°C for 3 h to initiate exfoliation and transfer process to the target substrate. After being transferred, the GaP layer is subjected to a secondary annealing process in a conventional furnace. The samples were annealed in vacuum at 600 °C for 12 h to promote crystallinity recovery and surface quality enhancement.

**AFM measurement**

Atomic Force Microscopy (AFM) measurements were performed using a Dimension Edge system (Bruker), operating in tapping mode to minimize tip–sample interactions and preserve surface integrity during imaging. High-resolution topographic data were acquired with Tap300Al-G silicon cantilevers, specifically engineered for tapping mode and featuring nominal resonance frequencies around 300 kHz and force constants optimized for nanoscale surface profiling. Scans were conducted over a 15 × 15 µm² area to capture mesoscale morphological features, with images recorded at a resolution of 1024 × 1024 pixels to ensure high-fidelity surface reconstruction. A scan rate of 0.5 Hz was selected to achieve an optimal trade-off between the acquisition speed, lateral resolution, and signal-to-noise ratio, enabling reliable analysis of fine surface features.

**Raman Mapping**

Raman mapping of the exfoliated flakes before and after annealing was conducted using a Renishaw inVia Raman microscope (Renishaw, UK) in a backscattering configuration equipped with a 532 nm excitation laser. The laser power at the sample surface was carefully adjusted to prevent sample heating and degradation. A 2400 lines/mm diffraction grating was employed to achieve high spectral resolution suitable for resolving fine vibrational features. Raman signals were collected through a 100× objective (NA = 0.9), resulting in a lateral resolution of approximately 1 µm. The mapping was performed in a point-by-point raster mode over a defined rectangular area of approximately 869.25 µm² (28.5 µm × 30.5 µm) with a step size of 0.5 µm, yielding a total of 3596 spectra. Each spectrum was acquired with an integration time of 16 s and a single accumulation, providing high signal-to-noise data while preserving acquisition throughput. Silicon at 520.7 $cm^{-1}$ was used as a calibration standard to ensure wavenumber accuracy across the dataset.

**Ellipsometry**

Spectroscopic ellipsometry measurements were performed using a SENTECH ellipsometer to investigate the optical properties and thickness of the GaP thin film. The system operated over a broad spectral range of 400–2200 nm, covering both the visible and near-infrared regimes, enabling accurate determination of dispersive optical constants. Measurements were conducted using a beam diameter of approximately 2 mm to ensure spatial averaging over a representative surface area, thereby minimizing the influence of local thickness variations or implantation-induced inhomogeneities. Ellipsometric data ($\Psi$ and $\Delta$) were acquired at multiple angles of incidence of 40°, 50°, and 60° to enhance the reliability and uniqueness of the optical model fitting and to reduce parameter cross-correlation during regression analysis. The polarizer was set at fixed orientations of +45° and −45°, enabling optimal sensitivity to polarization-dependent reflectance changes, while the analyzer was rotated in 8 steps. The ellipsometric data was fitted with the Cody-Lorentz dispersion model [26,27].

**Acknowledgements**

The authors acknowledge the Fraunhofer Attract Grant SILIQUA No. 40-04866, the BMFTR projects GOI-4-IQ-Nano Grant No. 13N17106, MEXSIQUO Grant No. 13N16967, and SINNER Grant No. 16KIS1792, and the Collaborative Research Center (CRC/SFB) 1375 NOA.

# Supplementary Information: Direct Wafer Bonding of Crystal-Ion-Sliced GaP Thin Films for Photonic Applications

Hossein Esfandiar [1,5,*], Emanuel Glück [1], Fabian Ganss [2], Ulrich Kentsch [2], Dennis Arslan[1], Jana Paeschke [1], Sebastian Ritter [1,5,6], Andreas Ihring [3], Muyi Yang [4,5,6], Isabelle Staude [4,5,6] Stefan Facsko [2], Falk Eilenberger [1,5,6], Carolin Rothhardt [1], Sebastian W. Schmitt [1,5,+]

[1] Fraunhofer-Institute for Applied Optics and Precision Engineering IOF, Albert-Einstein-Str. 7, 07745 Jena, Germany

[2] Institute of Ion Beam Physics and Materials Research, Helmholtz-Zentrum Dresden-Rossendorf, Bautzner Landstr. 400, 01328 Dresden, Germany

[3] Leibniz Institute of Photonic Technology, Albert-Einstein-Str. 9, 07745 Jena, Germany

[4] Institute of Solid-State Physics, Friedrich Schiller University, Max-Wien-Platz 1, 07743 Jena, Germany

[5] Institute of Applied Physics, Abbe Center of Photonics, Friedrich Schiller University, Albert-Einstein-Str. 15, 07745 Jena, Germany

[6] Max Planck School of Photonics, Albert-Einstein-Str. 15, 07745 Jena, Germany

*hossein.esfandiar@iof.fraunhofer.de

+sebastian.wolfgang.schmitt@iof.fraunhofer.de

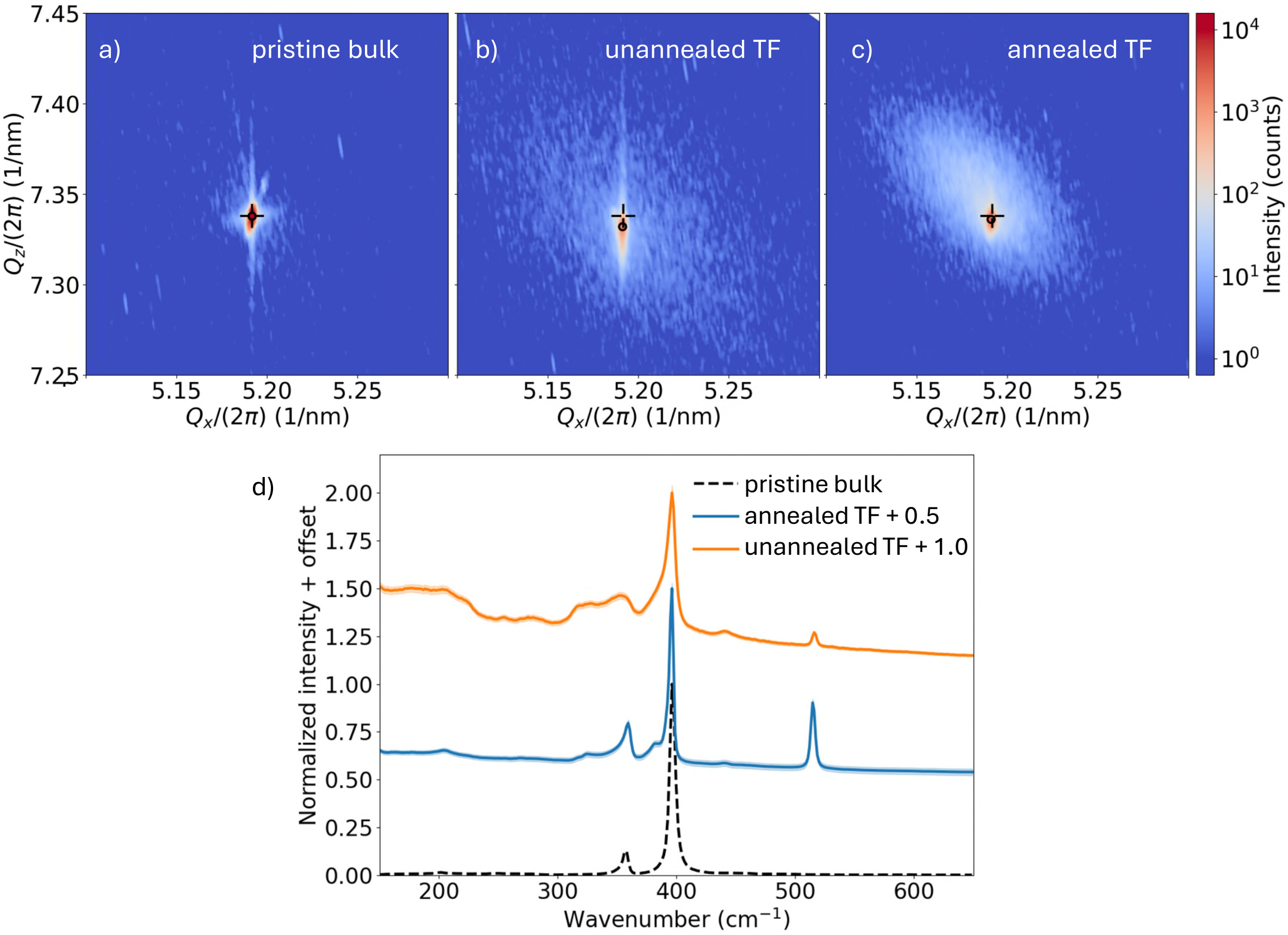

**Figure S1.** Structural investigation of the GaP before and after exfoliation, bonding and annealing. Reciprocal space maps (log intensity) of GaP around the (224) reflection for **(a)** pristine bulk GaP, **(b)** $He^+$ implanted, exfoliated and bonded GaP thin films, and **(c)** the same sample after annealing for 7 h 600°C in vacuum. **(d)** Raman spectra of bulk GaP, the $He^+$ ion sliced and bonded GaP film before annealing, and the film after annealing at 600 °C for 7 h in vacuum.

Figure S1 shows the (224) reciprocal space maps (RSMs) of the pristine bulk GaP (Fig. S1a), the GaP thin film following crystal $He^+$ ion slicing and bonding to a silica substrate (Fig. 3b), and the same bonded GaP thin film after annealing at 600 °C for 7 h in vacuum (Fig. 3c). In each map, the black open circle denotes the peak position obtained from a two-dimensional Gaussian fit of the data while the black cross indicates the reciprocal lattice coordinates of the bulk pristine GaP ($Q_x/(2\pi)$ = 5.1917 nm$^{-1}$, $Q_z/(2\pi)$ = 7.3380 nm$^{-1}$). Accordingly, the cross coincides with the fitted peak center in Fig. 3a, while the offsets between circle and cross observed in Figs. 3b and 3c indicate deviations from the pristine bulk GaP lattice.

The pristine bulk GaP crystal (Fig. 3a) exhibits a sharp, diffraction peak with minimal diffuse scattering, characteristic of a high-quality single crystal. The narrow-fitted widths of FWHM($Q_x$)/(2π) = 1.63×10$^{-3}$ nm$^{-1}$ and FWHM($Q_z$)/(2π) = 5.05×10$^{-3}$ nm$^{-1}$ confirm the excellent crystalline quality and low defect density of the pristine bulk material.

Following crystal ion slicing and direct bonding to silica (Fig. S1b), the diffraction peak becomes noticeably broader and is accompanied by pronounced diffuse scattering surrounding the Bragg reflection. The scattering is elongated primarily along the out-of-plane direction, indicating an increased distribution of lattice spacings associated with implantation-induced strain and structural disorder. The fitted peak shifts to ($Q_x$, $Q_z$)/(2π) = (5.1912, 7.3321) nm$^{-1}$. The significant reduction in $Q_z/(2\pi)$ indicates an expansion of the out-of-plane lattice parameter compared with pristine bulk GaP likely due to the presence of implanted ions but is less pronounced than for the $He^+$/ $H^+$ implanted film (Fig. 3b). The fitted peak widths increase to FWHM($Q_x$)/(2π) = 2.49 × 10$^{-3}$ nm$^{-1}$ and FWHM($Q_z$)/(2π) = 8.11 × 10$^{-3}$ nm$^{-1}$, demonstrating substantial implantation-induced strain broadening and increased lattice disorder introduced during the ion slicing and bonding processes.

After annealing at 600 °C for 7 h in vacuum (Fig. S1c), the diffraction peak moves back closer to the nominal bulk GaP reciprocal lattice position, with the fitted peak located at ($Q_x$, $Q_z$)/(2π) = (5.1910, 7.3362) nm$^{-1}$ indicating a significant recovery of the implantation-induced lattice distortion, particularly in the out-of-plane direction. Compared with the unannealed film, the diffuse scattering halo changes shape and the Bragg reflection appears more localized. The fitted peak widths decrease to FWHM($Q_x$)/(2π) = 2.73 × 10$^{-3}$ nm$^{-1}$ and FWHM($Q_z$)/(2π) = 6.88 × 10$^{-3}$ nm$^{-1}$. These values are close to the pristine bulk GaP, showing that thermal annealing substantially relieves strain and improves the crystalline quality of the transferred GaP layer.

Overall, the reciprocal space maps demonstrate that the ion slicing and bonding processes introduce lattice strain and structural disorder, whereas post-bonding annealing effectively relaxes much of the implantation-induced strain, bringing the transferred GaP thin film closer to the structural state of the bulk crystal.

Figure S1d presents the Raman spectra of pristine bulk GaP, the $He^+$ ion-sliced and directly bonded GaP thin film before annealing, and the transferred film after annealing at 600 °C for 7 h in vacuum. The pristine GaP spectrum exhibits the characteristic transverse optical (TO) and longitudinal optical (LO) phonon modes of crystalline GaP, with sharp and well-defined Raman peaks indicative of excellent crystalline quality. Pseudo-Voigt fitting yields TO and LO peak positions of 356.78 ± 0.58 cm$^{-1}$ and 396.14 cm$^{-1}$, with corresponding full widths at half maximum (FWHM) of 5.44 ± 2.18 cm$^{-1}$ and 5.63 ± 0.27 cm$^{-1}$, respectively.

Following $He^+$ crystal ion slicing and direct bonding (unannealed film), the Raman spectrum exhibits a substantially elevated background together with broadened phonon features, reflecting implantation-induced lattice disorder, strain inhomogeneity, and defect-assisted scattering. Although both TO and LO modes remain observable, the TO peak shifts to 354.58 ± 2.69 cm$^{-1}$ and broadens to 9.11 ± 7.38 cm$^{-1}$, while the LO linewidth increases to 8.84 ± 0.84 cm$^{-1}$.

After annealing at 600 °C for 7 h, the Raman spectrum shows a marked recovery of crystalline quality. The background scattering is substantially reduced, and both phonon modes become sharper and more prominent. The TO phonon shifts to 358.29 ± 0.72 cm$^{-1}$ with a reduced linewidth of 6.46 ± 1.98 cm$^{-1}$, while the LO phonon linewidth decreases dramatically to 3.88 ± 0.38 cm$^{-1}$, becoming even narrower than that of the pristine bulk crystal.

Overall, the Raman analysis demonstrates that $He^+$ crystal ion slicing and bonding introduces significant lattice disorder and phonon broadening, whereas post-bonding annealing effectively relieves implantation-induced damage, reduces local strain fluctuations, and restores the crystalline quality of the transferred GaP film. These observations are in excellent agreement with the reciprocal space mapping results, which likewise show reduced lattice strain and improved structural order after annealing.